\documentclass[12pt]{article}   
\usepackage{amsfonts}   
\usepackage{amsmath}   
\usepackage{multicol}   
\usepackage{epsfig}   
\usepackage{graphicx}  
\usepackage{feynarts}
\def\G{\Gamma}    
\def\a{\alpha}

\begin{document}
\begin{flushright}
YITP-SB-04-26\\   
IHES/P/04/23 \\
CERN-PH-TH/2004-083\\
SLAC-PUB-10442\\
MPP-2004-48\\   
\end{flushright}       
     
\vskip 0.4 truecm
\Large
\bf

\centerline{On the Landau Background Gauge Fixing}
\centerline{and the IR Properties of YM Green Functions}   

\normalsize
\rm

\vskip 0.9 truecm
\large

\centerline{Pietro A. Grassi$^{a}$\footnote{\sf pgrassi@insti.physics.sunysb.edu}, Tobias Hurth$^{b}$\footnote{\sf tobias.hurth@cern.ch}, 
Andrea Quadri$^{c}$\footnote{\sf quadri@mppmu.mpg.de}}
\small
\vskip 0.2 truecm
\begin{center}
$^{a}$ C.N. Yang Institute for Theoretical Physics,  
SUNY at Stony Brook\\ 
Stony Brook, NY 11794-3840, USA,  \\
Dipartimento di Scienze,
Universit\`a del Piemonte Orientale, \\
C.so Borsalino 54, I-15100 Alessandria, ITALY, and \\
IHES, Le Bois-Marie, 35,  \\ 
route de Chartres, F-91440 Bures-sur-Yvette, FRANCE. \\
\vskip 0.3 truecm
$^{b}$ CERN, Theory Division,  
CH-1211 Geneva 23, SWITZERLAND and \\  SLAC, Stanford University, 
Stanford, CA 94309, USA \\ 
\vskip 0.3 truecm
$^{c}$ Max-Planck-Institut f\"ur Physik\\
(Werner-Heisenberg-Institut)\\
F\"ohringer Ring, 6 - D80805 M\"unchen, GERMANY\\
\end{center}
\vskip 0.4  truecm
\normalsize
\bf
\centerline{Abstract}

\rm
\begin{quotation}
We analyse the complete algebraic structure of the 
background field method for Yang--Mills theory 
in the Landau gauge and show 
several structural simplifications within this approach. 
In particular we present a new way to study the IR behavior 
of Green functions in the Landau  gauge and show that there 
exists a unique Green function whose  IR behaviour controls 
the IR properties of the gluon and the  ghost propagators.
\end{quotation}

\newpage
   
%%%%%%%%%%%%%%%%%%%%%%%%%%%%%%%%%%%%%%%%%%%%%%%%%%%%%%%%%%%%%

\section{Introduction}   

The use of the background field method (BFM) \cite{BFM} 
is an outstanding technique to simplify the computation of 
Green functions in gauge theories. This assessment should   
be extended beyond perturbation theory as demonstrated in 
numerous  studies (see for example \cite{studies});
in particular, it was shown that the BFM is most suitable 
in quantum field 
theoretical  computations around classical solutions of 
the field equations \cite{classical}. 

We have, however, to point out that the complete algebraic 
structure of the BFM in gauge theories has been scarcely 
used in the applications. Nevertheless this structure, encoded in 
the BRST formalism, has led to several important results: 
the proof of the equivalence of the correlation functions between  
physical observables computed in the BFM and in the conventional 
formalism \cite{AGS}, important simplifications in the 
computations of radiative corrections for the SM \cite{amt}, 
some progresses in constructing effective charges and 
gauge-invariant quantities \cite{PT} and the extensions to open algebras and 
non-trivial manifolds \cite{ghq}.  	

In the present paper, we exploit the complete 
algebraic structure of the BFM for Yang--Mills theory
in the Landau gauge. We recall that one can prove 
some non-renormalization theorems for the ghost-gluon vertex 
and for the ghost two-point functions in the Landau gauge. 
These properties are 
consequences of an {\it integrated} functional equation -- {\it 
the antighost equation} -- which can be shown to hold
for the quantum effective action
in the Landau gauge \cite{blasi,book}. 
Moreover, as we will show here, 
by combining the properties 
of the Landau gauge with the BFM, one can derive  a more 
powerful {\it local} equation for the antighost fields.
Another example of a local antighost equation was presented 
in \cite{aae}.

In the case of the Landau gauge, the commutation relation between 
the integrated antighost equation and the Slavnov--Taylor identities 
(STI) (which implement the BRST symmetry at the functional level) 
yields the integrated Ward--Takahashi identities
(WTI) for the $SU(N)$ rigid symmetry. 
In the Landau background gauge, the local antighost equation leads to local 
WTI. The latter are functional equations for the gauge 
fixed effective action and they 
carry all the relevant  
information on the original gauge symmetry 
of the ungauged classical 
theory. In the present context, the role of the WTI partly 
supersedes that of STI, which 
essentially implement 
only the relation between the background fields and the quantum 
ones. 

The implementation of the BRST symmetry at the quantum 
level requires the introduction of certain sources (called in 
the following antifields) to study the renormalization 
of the composite operators emerging from the BRST 
variation of the fields.
In the same way, the BFM 
requires new sources to be coupled to the variation 
of the action under the BRST transformations 
of the background fields. 
Those sources are indeed sufficient to implement 
the full set of symmetries of the theory at the 
quantum level. Moreover, 
we also show that some of the Green functions obtained by differentiating 
the functional generators (for one-particle-irreducible (1-PI)  
or connected graphs) with 
respect to these sources can be used to 
study IR properties of the gluon and ghost propagator. This 
is an interesting feature that is not usually taken into account for 
practical applications. 

In the paper, after working out the algebraic 
structure of the Landau-background 
gauge fixing and its symmetries, we study some 
applications of the formalism. Here we list them with 
some comments.

\begin{itemize}

\item
During the last ten years, some effort has been devoted 
to study the infrared behaviour of the Green functions
of QCD. Understanding the infrared properties is intimately 
related to grasping some information on the non-perturbative 
regime of the theory in the realm of confined gluons and quarks. 

One of the most used and promising techniques is lattice QCD, where the 
natural regularization and infrared cut-off has led to fundamental 
simulations to very low energies. However, 
other techniques have been used, such as those 
based on the  renormalization group (RG) equations and the 
Schwinger--Dyson equations. 

Let us first comment on RG equations. 
The renewed interest in the Wilson 
renormalization group stimulated by a paper by Polchinski 
\cite{Polchinski:1983gv} and the vast literature that 
followed it (see for example 
\cite{becchi-lectures}--\cite{paw} 
and  references thereof and therein)  yielded new applications 
of these techniques to the study of low-energy effective actions. The aim 
was to see emerging from the RG evolution the signals of confinements 
or new phases in the strong regime of QCD. However, in order 
to look after non-perturbative effects, one has to solve the functional 
equations non-perturbatively. This amounts to providing  suitable 
truncations and relatively simple ans\"atze for 
the Green functions involved in the computations. In the 
Landau gauge, some of these ans\"atze can be explicitly implemented 
and justified. Here we would like to point out that 
in addition to the Landau gauge, 
the BFM might give further
 simplifications and insights. To show this, we present a simple 
application and we leave to future publications a real non-perturbative 
computation in the Landau BFM.

Recently it has been shown  in Ref.~\cite{bonini} that within the
Wilson RG formalism a regularization that preserves the background 
gauge invariance can be  found. 
This regularization preserves also all linear identities associated to 
ghost or antighost equations of motion, but it breaks the STI. However, 
we show that the number of breaking terms of STI 
in the background Landau gauge is highly reduced.
Moreover, it is easy to solve the fine-tuning problem
by computing those counterterms 
which are needed in order to reabsorb the ST breakings,
thanks to the special structure of the STI in the Landau
background gauge. 
In contrast to the usual technique with BFM in non-singular gauges, 
the gauge symmetry 
is encoded in the WTI. The present analysis can be extended to chiral gauge 
theories whenever the anomaly is cancelled.\\ 

\item 
The technique developed for solving the 
Schwinger--Dyson equations (SDE) in the Landau gauge \cite{Alkofer}  
has also been used in the past to 
understand the non-perturbative properties of the theory. This 
gauge is the most suitable to justify the truncations used
and also favored due to the non-renormalization 
theorems.  Recently,  the role of the Gribov horizon in 
the dynamical mass generation       
in euclidean Yang-Mills theories has been investigated in the context of        the Landau gauge in \cite{sorella}.   
Nevertheless adding the background field method would 
enhance even more simplifications. As an illustration of these phenomena 
we rederive the non-renormalization theorems for the Landau background 
gauge, and in particular we show how the WTI replace the STI 
to uncover the gauge properties of the Green functions. 
We point out that the relevant properties of the IR behaviour 
of gluon and  
ghost propagators are encoded in a single Green function, obtained by 
differentiating 
the generating functional with respect 
to certain classical sources. As an aside we show the 
relation between this Green function and the Kugo--Ojima criterion for 
confinement \cite{Kugo:gm}. Some recent developements where 
the Landau-Background can be used is the Maldestam's 
approximation to SDE \cite{malde}.\\ 

\item
In general, the STI are complicated 
non-linear  functional equations, which are
rather difficult to handle outside of perturbation theory\footnote{Here 
we assume that a perturbative definition of BRST symmetry is possible.}.
Instead of studying the STI, one can, in the BFM,  
substitute part of them 
with linear WTI identities for the background gauge invariance. This 
simplifies also the study of counterterms at higher orders. This is 
a drastic simplification for supersymmetric QCD where the Landau gauge 
has been used for explicit computations.\\

\item
Concerning the famous problem of Gribov copies in covariant 
gauges, the use of BFM has several advantages. We should 
recall that the Gribov copies are induced by the vanishing of the 
Faddeev--Popov determinant.  However, a convenient way to 
overcome these difficulties is to use the 
technique developed in \cite{Becchi},
where a suitable patching of the configuration space is 
used by changing the background field $\hat A_{\mu}$. 
The Landau gauge indeed has Gribov copies that may 
affect the computations at low energy. 
The use of BFM in the Landau gauge might prevent this 
problem also in numerical 
simulations where the background can be changed. In \cite{Becchi} the 
prescription to move from one patch to another is given. 
The practical implications will be discussed elsewhere. \\

\item
The structure of the BRST symmetry with background 
fields is rather suggestive. As is known, the BRST symmetry, 
the anti-BRST symmetry, the ghost equations and the antighost 
equation combine in a symmetric fashion; it can be shown
that the commutation relation of  
the Nakanishi-Lautrup equation for the Lagrangian multiplier 
and   the linearized ST operator or the linearized anti-ST 
operator leads to the ghost and antighost equations, respectively. 
Further commutation relations between the linearized ST operator 
and the linearized anti-ST operator lead to the WTI.
\end{itemize}

The paper is organized as follows: in Section~\ref{sec:landau_gf} 
we present the Landau background 
gauge fixing, the STI and WTI for two-point functions and 
the non-renormalization theorem 
for the ghost-gluon-ghost vertex. 
In Section~\ref{sec:appl} we study the following applications: 
RG equations and the BFM, the IR behaviour of Green functions, 
the relation 
with Green functions with external classical sources, and 
the Kugo--Ojima criterion in the present framework.
We also comment on some details of the computations. 
Finally conclusions are presented in Section~\ref{sec:concl}.

%%%%%%%%%%%%%%%%%%%%%%%%%%%%%%%%%%%%%%%
     
\section{Landau Background Gauge Fixing}\label{sec:landau_gf}

We consider $SU(N)$ Yang--Mills theory and adopt the 
following conventions: $[T^{a}, T^{b}] = i f^{abc} T^{c}$, where 
$T^{a}$ are hermitian generators of the corresponding
Lie algebra and $f^{abc}$ are the associated real structure constants. The 
covariant derivative on adjoint fields $\Phi^{a}$ is 
given by $(\nabla_{\mu} \Phi)^{a} = \partial_{\mu} \Phi^{\a} + 
f^{abc} A_{\mu}^{b} \Phi^{c} = (\partial_{\mu} \Phi - i [A_{\mu}, \Phi])^{a}$ where 
$A_{\mu} = A^{a}_{\mu} T^{a}$ and $\Phi =\Phi^{a} T^{a}$. The $T^{a}$ 
are normalized in such a way that ${\rm tr}(T^{a} T^{b}) = {1\over 2} \delta^{ab}$. 

We impose the Landau background 
gauge fixing 
\begin{equation}\label{lb}
\hat\nabla_{\mu} ( A - \hat A)^{\mu} = 0\,,
\end{equation} 
where 
$\hat \nabla_{\mu}$ is the background covariant derivative, $A_\mu$ is the 
gauge field, and $\hat A_\mu$ is the background gauge field. The complete 
Lagrangian is obtained by the usual BRST prescription:
\begin{eqnarray}\label{lag}
{\cal L} &=& - {1\over 4}{\rm tr} F^{2}+ s \, {\rm tr}
\left( \bar c  \hat\nabla_{\mu} ( A - \hat A)^{\mu}  - A^{*} _{\mu} A^{\mu } + c^{*} c \right) =  \nonumber \\
& = &  {\rm tr} \left[-  {1\over 4}  F^{2} + b \hat\nabla_{\mu} ( A - \hat A)^{\mu} - 
\bar c \left( \hat \nabla^{\mu} \nabla_{\mu} c  - \nabla_{\mu} \Omega^{\mu} \right) \right . \nonumber \\
& & \left . ~~~~ + A^{*} _{\mu} \nabla^{\mu} c + i c^{*} c^{2} \right],
\end{eqnarray}
where $s$ is the generator of the BRST transformations: 
\begin{eqnarray}\label{lbII}
&&s\, A_{\mu } =  \nabla^{\mu} c\,, ~~~~~ 
s\, c =  i c^{2}\,, ~~~~~ 
s\, \hat A_{\mu} = \Omega_{\mu}\,,  ~~~~~ 
s\, \Omega_{\mu} = 0\,,  \nonumber \\
&&s\, \bar c = b\,, ~~~~~~~~~~~
s\, b = 0\,, ~~~~~~~
s\, A^{*}_{\mu} =0\,, ~~~~~~~
s\, c^{*} =0\,.
\end{eqnarray} 

The dimensions of the fields and antifields 
and their ghost number are summarized in Table \ref{tab1}.
\begin{table}
\begin{center}
\begin{tabular}{|c|c|c|c|c|c|c|c|c|}
\hline
        & $A_\mu$ & $b_\mu$ & $\bar c$ & $c$ & $\hat A$ & $\Omega$ & $A_\mu^*$ & $c^*$ \cr
\hline
Dim.   & $1$     &     $2$ &      $2$ & $0$ &      $1$ &      $1$ &       $3$ & $4$   \cr
\hline
Gh. numb. & $0$     &     $0$ &     $-1$ & $1$ &      $0$ &      $1$ &      $-1$ & $-2$ \cr
\hline
\end{tabular}
\end{center}
\caption{Ghost number and dimension of the fields and the antifields\label{tab1}}
\end{table}

At the classical level the theory is characterized by the 
STI for the action $S = \int d^4x \, {\cal L}$,

\begin{eqnarray}\label{STI}
{\cal S}(S) = \int d^{4}x \, {\rm tr} \left( {\delta S \over \delta A^{*}_{\mu} } {\delta S \over \delta A_{\mu} } 
+ {\delta S \over \delta c^{*} }{\delta S \over \delta c }  + b {\delta S \over \delta \bar c } + \
\Omega_{\mu} {\delta S \over \delta \hat A_{\mu} }\right) = 0\,,
\end{eqnarray}
by the equation for the Nakanishi--Lautrup field $b$
\begin{equation}\label{NLE}
{\delta S \over \delta b} = \hat\nabla_{\mu} ( A - \hat A)^{\mu}
\end{equation}
and by the antighost equation
\begin{equation}\label{AGE}
{\delta S \over \delta c} - \hat\nabla_{\mu} {\delta S \over \delta \Omega_{\mu}} 
+ i \left[ \bar c, {\delta S \over \delta b}\right]
- \nabla_{\mu} A^{* \mu} + i [c^{*}, c]  = 0 \,.
\end{equation}
The last equation is a consequence of the 
special gauge fixing we have adopted.

Due to the special choice of the background gauge fixing,
 all the composite operators obtained by differentiating the action $S$ 
with respect to the 
ghost field $c$ can be re-expressed in terms of functional derivatives of the 
action with respect to the source $\Omega_{\mu}$ and with respect to $b$. 
This can be shown by using integration by parts.  
A fundamental property to be used is 
$[\hat\nabla_{\mu}, \nabla^{\mu}] = - i \hat \nabla_{\mu} (A -\hat A)^{\mu}$. 

Equation (\ref{AGE})  tells us that 
the ghost field $c$ will not get an independent renormalization 
and that, in addition, the 
dynamics of $c$ is completely 
fixed by this equation. Notice that in the case of the Landau gauge fixing without background fields only an integrated version of (\ref{AGE}) was derived \cite{blasi}.

As a consequence 
of equations.~(\ref{STI}) -- (\ref{AGE}) we can obtain the following two functional equations:
the Faddeev--Popov equation
\begin{eqnarray}\label{GE}
{\cal S}_{S}\left( {\delta S \over \delta b} - \hat\nabla_{\mu} ( A - \hat A)^{\mu}
\right) - {\delta \over \delta b} {\cal S}(S) = {\cal G}(S) \equiv
{\delta S \over \delta \bar c} +  \hat\nabla_{\mu} {\delta S \over \delta A^{*}_{\mu}} 
- \nabla_{\mu} \Omega^{\mu} = 0\,,
\end{eqnarray}
where ${\cal S}_{S}$ is the linearized ST operator,
defined by
\begin{eqnarray}\label{lin_ST_op}
{\cal S}_S = \int d^4x \, {\rm tr} \Big (
\frac{\delta S}{\delta A^*_\mu} \frac{\delta}{\delta A_\mu}
+
\frac{\delta S}{\delta A_\mu} \frac{\delta}{\delta A_\mu^*}
+
\frac{\delta S}{\delta c^*} \frac{\delta}{\delta c}
+
\frac{\delta S}{\delta c} \frac{\delta}{\delta c^*}
+ b \frac{\delta}{\delta \bar c}
+ \Omega^\mu \frac{\delta}{\delta \hat A_\mu}
\Big ) \, ,
\end{eqnarray}
 and the WTI 
\begin{eqnarray}\label{WTI}
&&
{\cal S}_{S}\left( {\delta S \over \delta c} - \hat\nabla_{\mu} {\delta S \over \delta \Omega_{\mu}} 
- \nabla_{\mu} A^{* \mu} +i  [c^{*}, c] +i [ \bar c, {\delta S \over \delta b}] \right) 
+ {\delta \over \delta c} {\cal S}(S) \equiv {\cal W}(S) = \nonumber \\
&&  - \nabla_{\mu} {\delta S \over \delta A_{\mu}}  - 
\hat \nabla_{\mu} {\delta S \over \delta \hat A_{\mu}}  + 
[A^{*}_{\mu}, {\delta S \over \delta A^{*}_{\mu}} ] + 
[c^{*}, {\delta S \over \delta c^{*}} ] + 
[c, {\delta S \over \delta c} ] + 
[\bar c, {\delta S \over \delta \bar c} ] +  
\nonumber \\
&& 
+
[\Omega_{\mu}, {\delta S \over \delta \Omega_{\mu}} ] + 
[b, {\delta S \over \delta b} ] = 0\,. 
\end{eqnarray}

In the case of 
ordinary 
Landau gauge fixing (without the background), the anticommutation relation 
between the antighost equation and the STI yields the integrated WTI.
On the other hand in the Landau background gauge 
the local equation for the ghost field (\ref{AGE}) leads directly 
to the local WTI for the background gauge invariance. 

Moving to the quantum theory and assuming a renormalization 
prescription that  preserves the identities (\ref{STI}) -- (\ref{WTI}), we substitute 
the tree-level action $S$ with the 1-PI generating functional $\Gamma$.

In order to simplify the analysis of the 1-PI Green functions in the following 
sections, we introduce the new functional $\hat \Gamma$
\begin{equation}\label{hatfunc}
\hat \Gamma = \Gamma - 
\int d^{4}x \, {\rm tr} \left( b \hat\nabla_{\mu} (A -\hat A)^{\mu} + 
\Omega^{\mu} \nabla_{\mu} \bar c
\right)\,.
\end{equation}
The introduction of $\hat \Gamma$
 simplifies equation (\ref{NLE}), which 
in particular implies $\delta \hat\Gamma / \delta b =0$;  
it also simplifies equation  (\ref{GE}). It modifies the ST identities in (\ref{STI}),
but it is easy to see that they have the same 
form,  provided that we substitute $A^{*}_{\mu}$ with 
$\hat A^{*}_{\mu} = 
A^{*}_{\mu} + \hat \nabla_{\mu} \bar c$. The WTI in (\ref{WTI}) are 
preserved from  all manipulations, and the ghost equation (\ref{GE}) 
is automatically solved. Finally, it is convenient to introduce 
$\hat\Omega_{\mu} = \Omega_{\mu} + \hat\nabla_{\mu} c$, 
in order to simplify the antighost equation (\ref{AGE}). In particular, the 
latter tells us that the only dependence of the functional 
$\hat \Gamma[ A, \hat A,\hat A^{*}, \hat \Omega, c]$ 
upon the ghost field $c$ is  given by 
\begin{equation}\label{fundep }
\frac{\delta \hat \Gamma}{\delta c} = 
\nabla^{\mu} \hat A^{*}_{\mu} -i [c^{*}, c]\,,
\end{equation}
and therefore, we can finally introduce the functional 
\begin{equation}\label{finaaction}
\hat \Gamma' = \hat\Gamma - \int d^{4}x 
\, {\rm tr}\left( \hat A^{*}_{\mu} \nabla^\mu c + i 
c^{*} c^{2} \right)\,,
\end{equation}
which removes the explicit dependence upon $c$;
$\hat \G'$ is a functional of $A_\mu, \hat A_\mu, c^*$ and $\hat \Omega_\mu$
only.
 The complete 
dynamics is constrained by the remaining two functional equations:  
\begin{eqnarray}\label{finaleqs}
&&{\cal S}(\hat \Gamma') = 
\int d^{4}x \, 
{\rm tr} \left( {\delta \hat\Gamma' \over \delta \hat A^{*}_{\mu} } 
{\delta \hat\Gamma' \over \delta A_{\mu} } +
\hat\Omega_{\mu} {\delta \hat\Gamma' \over \delta \hat A_{\mu} }\right) = 0\,,  \end{eqnarray}
\begin{eqnarray}\label{final_WTI}
&& \!\!\!\!\!\!\!\!\!\!\!\! \!\!\!\!\!\!\!\!\!\!\!\! {\cal W}(\hat \Gamma') = 
 - \nabla_{\mu} {\delta \hat\Gamma' \over \delta A_{\mu}}  - 
\hat \nabla_{\mu} {\delta \hat\Gamma' \over \delta \hat A_{\mu}}  + 
[\hat A^{*}_{\mu}, {\delta \hat\Gamma' \over \delta \hat A^{*}_{\mu}} ] 
+ [c^{*}, {\delta \hat\Gamma'  \over \delta c^{*}} ] + 
[\hat\Omega_{\mu}, {\delta \hat\Gamma'  \over \delta \hat\Omega_{\mu}} ] 
= 0\,.
\end{eqnarray}

The structure of the gauge group is completely encoded in the 
WTI. The dynamics of the ghost fields is encoded in the 
functional dependence of $\hat A^{*}_{\mu}$ and $\hat \Omega_{\mu}$; 
therefore, the only Green functions with external fields with 
non-vanishing ghost number have external 
$\hat A^{*}_{\mu}, \hat \Omega_{\mu}$ or $c^{*}$ fields. This is an enormous 
simplification, because these three fields are classical sources and 
not quantum fields. In order to clarify this point further, we 
can observe the following: non-trivial relations between 
Green functions can be obtained from (\ref{STI}) by differentiation 
with respect to a field carrying positive ghost number. However, the 
only positive charged field is $\hat \Omega$, so that  we have to 
take at least one derivative with respect to this field. This yields 
a relation 
among Green functions with one background field and those Green 
functions with only quantum gauge fields. So, the only information encoded 
in (\ref{STI}) is the relation between background Green functions $\hat\Gamma'_{\hat A \dots \hat A}$ 
and those with quantum fields $\hat \Gamma'_{A \dots A}$. 
The rest of the information on the 
symmetry of the theory is encoded in equations (\ref{WTI}) and 
(\ref{final_WTI}), 
which are linear identities and thus simpler than equation (\ref{STI}).

\medskip
Before ending this section, we just want to recall that
the existence of the anti-ghost equation in the 
Landau background gauge can be detected in a different 
way. As is well-known,  the BRST symmetry can be 
paired with an additional symmetry, the anti-BRST symmetry,  
where the ghost $c$ and the antighost $\bar c$ are 
exchanged. More precisely, one defines the anti-BRST differential $\bar s$
by
\begin{eqnarray}
\bar s A_\mu = \nabla_\mu \bar c \, , ~~~~ \bar s \bar c = i \bar c^2 \, 
~~~~ \bar s c = - b - i [c,\bar c] \, , ~~~~ \bar s b = - i [b,\bar c] \, .
\label{antiBRST_1}
\end{eqnarray}
We will not enter in all details here (but a similar 
analysis has been made in \cite{Baulieu:2000xi}). 
We only point out that besides the STI 
operator ${\cal S}_{S}$ (see (\ref{lin_ST_op})) one can correspondingly
introduce
the paired $\bar{\cal S}_{S}$ that generates the 
anti-BRST transformations. The two BRST 
operators anticommute and are both nilpotent. 

One then finds,  in complete analogy with equation (\ref{GE}): 
\begin{eqnarray}
\label{BBI}
%&&
%{\cal S}_{\Gamma}\left( {\delta \Gamma \over \delta b} - \hat\nabla_{\mu} ( A - \hat A)^{\mu}
%\right) - {\delta \over \delta b} {\cal S}(\Gamma) = 
%{\cal G} \Gamma = 0\,,  \nonumber \\
%&&
\bar{\cal S}_{S}
\left( {\delta S \over \delta b} - \hat\nabla_{\mu} ( A - \hat A)^{\mu}
\right) - {\delta \over \delta b} \bar{\cal S}(S) = 
\bar{\cal G} \Gamma = 0
\end{eqnarray}
where %${\cal G}$ and 
$\bar{\cal G}$ is the differential functional operator
appearing in %the ghost and 
the antighost equation %(\ref{GE}) and 
(\ref{AGE}). 
Hence in the presence of the anti-BRST symmetry the structure of the 
ghost and antighost equations are direct consequences of the choice 
of gauge fixing. 
Furthermore, it is possible to recast the gauge fixing in a  
BRST--anti-BRST exact expression. 
Since the details of 
the derivation of such a result are rather trivial and common in 
the literature we do not dwell on this point any further.  
We just 
observe that the field $\Omega_{\mu}$ is the source of
$\nabla_{\mu} \bar c$. The latter is the anti-BRST transformation 
of the gauge field $A_{\mu}$. Thus $\Omega_{\mu}$ is the antifield 
of $A_{\mu}$ with respect to the anti-BRST symmetry. Moreover, it 
is natural to view $A^{*}_{\mu}$ as the anti-BRST transformation 
of the background field $\hat A_{\mu}$. 
For additional details we refer the reader to Ref.~\cite{Baulieu:2000xi}.
  
%%%%%%%%%%%%%%%%%%%%%%%%%%%%%%%%%%%%%%

\subsection{Identities for two-point functions}
 
In this subsection,  
we use the Green functions generated by $\Gamma$ without the 
 redefinition discussed above.
This leads to relations between the 
 scalar functions appearing in the tensor decomposition of each single 
 Green function. Those can be computed directly by Feynman diagrams 
 stemming from the action (\ref{lag}). 
 
 Let us start from the ghost sector. There we have 
 four two-point functions: $\Gamma_{c\bar c}, \Gamma_{c A^{*}_{\nu}}, 
 \Gamma_{\Omega_{\mu} \bar c}$ and $\Gamma_{\Omega_{\mu} A^{*}_{\nu}}$. 
 By power-counting they are all divergent quantities, but 
 they are related by means of (\ref{GE}) and (\ref{AGE})
 \begin{eqnarray}\label{I}
\Gamma_{c\bar c}(p) & = & i p_{\nu} \Gamma_{c A^{*}_{\nu}}(p) \,, \nonumber  \\
\Gamma_{\Omega_{\mu} \bar c}(p) & = & 
i p_{\nu} \Gamma_{\Omega_{\mu} A^{*}_{\nu}}(p) +i p^{\mu} \,, \nonumber  \\
\Gamma_{\bar c c}(p) & = & 
-i p_{\mu} \Gamma_{\bar c \Omega_{\mu}}(p)  \,,
\nonumber  \\
\Gamma_{A^{*}_{\nu} c}(p) & = & -i p_{\mu} \Gamma_{A^{*}_{\nu} \Omega_\mu}(p) +i p^{\nu} \,.
\end{eqnarray}
We decompose the above Green functions in terms of scalar 
invariants in the following way (we omit the colour indices unless  
they are strictly necessary):  
\begin{eqnarray}\label{II}
 \Gamma_{c\bar c}(p) &=& - p^{2} G^{-1}(p) \,,\nonumber \\
 \Gamma_{A^{*}_{\nu}c}(p) &=& i p^{\nu}  A(p) \,,\nonumber \\
 \Gamma_{\Omega_{\mu} \bar c}(p) &=& i p^{\mu} B(p)\,, \nonumber \\
 \Gamma_{ A^{*}_{\nu} \Omega_{\mu}}(p) &=& - (L^{\mu\nu} C^{L}(p) + 
 T^{\mu\nu} C^{T}(p))\,,
\end{eqnarray}
where $L^{\mu\nu} = p^{\mu}p^{\nu}  /p^{2}$ and $T^{\mu\nu} = 
(g^{\mu\nu} - p^{\mu} p^{\nu} /p^{2})$.
% and the $C^{T}$ differs from 
%$C^{L}$ only for non-local terms. 
%{[\bf This sentence should be further clarified.]}
The divergent part of $C^{T}$ is 
equal to the divergent part of $C^{L}$. 
Equations (\ref{I}) yield 
\begin{equation}
\label{ III}
A(p) = B(p) = G^{-1}(p) = C^{L}(p) +1\,, 
\end{equation}
while $C^{T}(p)$ remains unconstrained. 
 
In the bosonic sector we have to 
 study the Green functions 
 $$
 \Gamma_{bb}(p)\,, 
 ~\Gamma_{b A_{\mu}}(p)\,, 
 ~\Gamma_{A_{\mu} A_{\nu}}(p)\,, 
 ~\Gamma_{b \hat A_{\mu}}(p)\,, 
 ~\Gamma_{\hat A_{\mu} A_{\nu}}(p)\,, 
 ~\Gamma_{A_{\mu} A_{\nu}}(p)\,.
 $$
 From the Nakanishi--Lautrup equation (\ref{NLE}) we 
 get
 \begin{equation}
\label{IV }
 \Gamma_{bb}(p) =0\,, ~~~~
 \Gamma_{A^{b}_{\mu} b^a}(p) = i \delta^{ab} \, p^{\mu}\,, ~~~~
 \Gamma_{\hat A^{b}_{\mu} b^a}(p) = - i \delta^{ab} p^{\mu}\,, 
\end{equation}
to all orders. For the two-point functions with gauge fields and their 
backgrounds,  we adopt the following decomposition
\begin{eqnarray}\label{V}
\Gamma_{A_{\mu} A_{\nu}}(p) 
&=& L^{\mu\nu} E^{L}(p) + T^{\mu\nu} E^{T}(p) \,, \nonumber \\
\Gamma_{\hat A_{\mu} A_{\nu}}(p) 
&=& L^{\mu\nu} F^{L}(p) + T^{\mu\nu} F^{T}(p) \,, \nonumber \\
\Gamma_{\hat A_{\mu} \hat A_{\nu}}(p) 
&=& L^{\mu\nu} H^{L}(p) + T^{\mu\nu} H^{T}(p) \,.
\end{eqnarray}
They have to satisfy the STI
 \begin{eqnarray}
&& \Gamma_{\Omega_{\nu} A^{*}_{\rho}}(p) 
\Gamma_{A_{\rho} A_{\mu}} (p)  +  
\Gamma_{\hat A_{\nu} A_{\mu}} (p) = 0 \, , 
 \\
&& \Gamma_{\Omega_{\nu} A^{*}_{\rho}} (p)
\Gamma_{A_{\rho} \hat A_{\mu}} (p) + 
\Gamma_{\hat A_{\nu} \hat A_{\mu}} (p) = 0 \, ,
\end{eqnarray}
and the WTI 
\begin{eqnarray}
p_{\mu}  \Gamma_{A_{\mu} A_{\nu}}(p)  + 
p_{\mu}  \Gamma_{\hat A_{\mu} A_{\nu}}(p) =0\,, \nonumber \\
p_{\mu}  \Gamma_{\hat A_{\mu} A_{\nu}}(p)  + 
p_{\mu}  \Gamma_{\hat A_{\mu} \hat A_{\nu}}(p)  =0\,,
\end{eqnarray} 
 leading to the equations
 \begin{eqnarray}
&& H^{L}  = -F^{L} =  E^{L} \,,  \nonumber \\
&& H^{L}  = - C^{L} F^{L}\,,  ~~~~~ F^{L}  = - C^{L} E^{L}  \nonumber \\
&& H^{T}  = - C^{T} F^{T}\,,  ~~~~~ F^{T}  = - C^{T} E^{T} \,. 
\label{eq_ward_st}
\end{eqnarray}
From the Nakanishi--Lautrup equation (\ref{NLE}), there follows
the transversality of
the gluon propagator $(\G_{A_\mu A_\nu})^{-1}$. Hence
$E^L=0$ and therefore from the first of equations (\ref{eq_ward_st}) 
$H^{L} = F^{L}=E^{L} =0$. Moreover from equations (\ref{eq_ward_st}) 
we get
$$
H^{T} = (C^{T})^{2} E^{T}\,.
$$
%
%{\bf Comment - From equations (\ref{eq_ward_st}) alone one cannot conclude that
%$H^{L} = F^{L}=E^{L} =0$. The Nakanishi-Lautrup equation is also needed.}
%
 To conclude,  we note that all relevant Green functions are 
 completely fixed in terms of three scalar invariants 
 $C^{T}, E^{T}$ and $C^{L}$. The last one is given 
 in terms of the two-point function of the ghost field and can be 
 computed from the  corresponding SDE. 
 The two-point function $E^{T}$ is the 
 transverse contribution to the gluon propagator 
 and can also be computed by the corresponding SDE; finally, the 
 scalar function $C^{T}$  can be computed by the two-point 
 function of the background propagator. However, both $C^{L}$ and 
 $C^{T}$ are the scalar functions appearing in
 the Green function $\Gamma_{\Omega_{\mu} A^{*}_{\nu}}(p)$;  
  they might therefore  be computed non-perturbatively 
 by solving the corresponding SDE (which we are going to 
 discuss in the next section) or from the RGE. 
 
%%%%%%%%%%%%%%%%%%%%%%%%%%%%%%%%%%%%

\subsection{Non-renormalization of the ghost--gluon vertex 
 $\Gamma_{c \bar c A_{\mu}}$}

In order to illustrate our method, we derive the non-renormalization of the 
gluon-ghost vertex and other useful relations between  Green 
functions in the same sector. 
We are interested in the following functions\footnote{We use the 
notation $\Gamma_{\Phi_{1} \dots \Phi_{n-1} \Phi_{n}}(p_{2}, \dots p_{n-1}, p_{n})$ where $p_{i}$ is the ingoing momentum of the field 
$\Phi_{i}$. For the field $\Phi_{1}$ we always assume the momentum 
conservation $p_{1} = - \sum_{i\neq 1} p_{i}$ and we omit it in the 
notation.}: 
$$
\Gamma_{c \bar c A_{\mu}}(p,q)\,,
~\Gamma_{\Omega_{\nu} \bar c A_{\mu}}(p,q)\,, 
~\Gamma_{c A^{*}_{\rho} A_{\mu}}(p,q)\,, 
~\Gamma_{\Omega_{\nu} A^{*}_{\rho} A_{\mu}}(p,q)\,.
$$
They satisfy the identities
\begin{eqnarray}\label{VI}
\Gamma_{c^a \bar c^b A_{\mu}^c}(p,q) &=& 
- i (p+q)_{\nu} \Gamma_{\Omega^a_{\nu} \bar c^b A^c_{\mu}}(p,q) + 
f^{abd} \Gamma_{b^d A^c_{\mu}}(q) \,, \nonumber \\
\Gamma_{c \bar c A_{\mu}}(p,q) &=& 
i p_{\rho} \Gamma_{c A^{*}_{\rho} A_{\mu}}(p,q) \nonumber \\
\Gamma_{\Omega^{a}_{\nu} \bar c^{b} A^{c}_{\mu}}(p,q) &=& 
i p_{\rho} \Gamma_{\Omega^{a}_{\nu} A^{*,b}_{\rho} A^{c}_{\mu}}(p,q) + 
f^{abc} g_{\mu\nu} \,, \nonumber \\
\Gamma_{c^{a} A^{*,b}_{\nu} A^{c}_{\mu}}(p,q) &=& 
-i (p+q)_{\rho} \Gamma_{\Omega^{a}_{\rho} A^{*,b}_{\nu} 
A^{c}_{\mu}}(p,q) + 
f^{abc} g_{\mu\nu} \,.
\end{eqnarray}
By power counting the Green function 
$\Gamma_{\Omega^{a}_{\nu} A^{*,b}_{\rho} A_\mu^c}(p,q)$ 
is finite (we recall that $\Omega_{\mu}^{a}$ has dimension 1, 
the gauge field $A_{\nu}^{a}$ has dimension 1 and 
the antifield $A^{*,a}_{\rho}$ has dimension 3). 
By equation (\ref{VI}) this implies that 
also the Green functions $\Gamma_{\Omega_{\nu} \bar c A_{\mu}}(p,q)$ and 
$\Gamma_{c A^{*}_{\rho} A_{\mu}}(p,q)$ are finite. Hence
we conclude that the ghost-gluon vertex 
$\Gamma_{c \bar c A_{\mu}}(p,q)$ is also finite. 
This is the well-known 
non-renormalization property of the ghost-gluon vertex in the Landau 
gauge. 

In addition, we can also 
prove that the ghost-background gluon vertex is finite
in the Landau background gauge. This amounts to 
deriving  the above equations for the Green functions 
$$
\Gamma_{c \bar c \hat A_{\mu}}(p,q)\,,
~\Gamma_{\Omega_{\nu} \bar c \hat A_{\mu}}(p,q)\,, 
~\Gamma_{c A^{*}_{\rho} \hat A_{\mu}}(p,q)\,, 
~\Gamma_{\Omega_{\nu} A^{*}_{\rho} \hat A_{\mu}}(p,q)\,.
 $$
and repeating the same argument. 

Furthermore, 
some useful identities are worth mentioning among the 
background and quantum Green functions. 
Some of them are obtained from the STI (\ref{STI}) and are given
(in condensed notation) by
\begin{eqnarray}\label{VII}
\Gamma_{c \bar c \hat A_{\mu}} & = & 
- \Gamma_{c A^{*}_{\rho}} \Gamma_{A_{\rho} \bar c \Omega_{\mu}} - 
\Gamma_{\Omega_{\mu} A^{*}_{\rho}} \Gamma_{A_{\rho} c \bar c} 
 + \G_{c \Omega_\mu c^*} \G_{c \bar c} 
\,, \nonumber \\
\Gamma_{c A^{*}_{\nu} \hat A_{\mu}} & = & 
- \Gamma_{c A^{*}_{\rho}} \Gamma_{A_{\rho} A^{*}_{\nu} \Omega_{\mu}} 
-\Gamma_{\Omega_{\mu} A^{*}_{\rho}} 
\Gamma_{A_{\rho} c A^{*}_{\nu}} 
+ \G_{c \Omega_\mu c^*} \G_{A^*_\nu c} 
\,, \nonumber \\
\Gamma_{\Omega_{\nu} \bar c \hat A_{\mu}} & = & -
\Gamma_{\Omega_{[\nu} 
A^{*}_{\rho}} \Gamma_{A_{\rho} \bar c \Omega_{\mu]}} 
+ \G_{\Omega_\nu \Omega_\mu c^*} \G_{\bar c c} \, ,
 \nonumber \\
\Gamma_{\Omega_{\nu} A^{*}_{\sigma} \hat A_{\mu}} & = & -
\Gamma_{\Omega_{[\nu} 
A^{*}_{\rho}} \Gamma_{A_{\rho} A^{*}_{\sigma}\Omega_{\mu]}} 
+ \G_{\Omega_\nu \Omega_\mu c^*} \G_{A^*_\sigma c} 
\end{eqnarray}
and 
from the WTI (\ref{WTI}) with the following form (again in condensed
notation):
\begin{eqnarray}\label{VIII}
&&-i(q+p)_{\mu} (\Gamma_{A_{\mu} c \bar c} + 
\Gamma_{\hat A_{\mu} c \bar c}) + 
\left(\Gamma_{c\bar c}(p) - \Gamma_{c \bar c}(q)\right) = 0\,, \nonumber \\
&&-i(q+p)_{\mu} (\Gamma_{A_{\mu} \Omega_{\nu} \bar c} + 
\Gamma_{\hat A_{\mu} \Omega_{\nu} \bar c}) + 
\left(\Gamma_{\Omega_{\nu}\bar c}(p) - \Gamma_{\Omega_{\nu} 
\bar c}(q)\right) = 0\,, \nonumber \\
&&-i(q+p)_{\mu} (\Gamma_{A_{\mu} c A^{*}_{\nu}} + 
\Gamma_{\hat A_{\mu} c A^{*}_{\nu}}) + 
\left(\Gamma_{c A^{*}_{\nu}}(p) - 
\Gamma_{c A^{*}_{\nu}}(q)\right) = 0 \, , \nonumber \\
&&-i(q+p)_{\mu} (\Gamma_{A_{\mu} \Omega_{\rho} A^{*}_{\nu}} + 
\Gamma_{\hat A_{\mu} \Omega_{\rho} A^{*}_{\nu}}) + 
\left(\Gamma_{\Omega_{\rho} A^{*}_{\nu}}(p) - 
\Gamma_{\Omega_{\rho} A^{*}_{\nu}}(q)\right) = 0\,.
\end{eqnarray}
The last terms in the R.H.S. of equation (\ref{VII}) are given
by superficially convergent functions involving one $c^*$ 
and at least one $\Omega_\mu$ times two-point functions
$\G_{\bar c c}$ or $\G_{A_\nu^* c}$. 
These two-point functions are in turn characterized by the parametrization
given in equation (\ref{II}), fulfilling the constraints in equation
 (\ref{ III}).

These equations show that the trilinear background 
Green functions with the insertion of an antighost or
an antifield $A^*$ are fixed in terms of the quantum Green 
functions (and vice versa). In particular the above relations
constrain the form of the ghost-gluon 
vertex.

%%%%%%%%%%%%%%%%%%%%%%%%%%%%%%%%%%%%%

\section{Applications}\label{sec:appl}

In this section we review some applications of the Landau gauge fixing
in the study of the IR properties of Yang--Mills theory and 
discuss the simplifications arising from the background Landau gauge
and its local antighost equation on some specific issues.

%%%%%%%%%%%%%%%%%%%%%%%%%%%%%%%%%%%%

\subsection{Renormalization Group Technique and BFM}

Recently, Bonini et al. discussed in \cite{bonini} the 
application of the BFM to the Wilson renormalization 
group technique. We can list two 
important advantages of that application: {\it i)} the WTI 
(\ref{WTI}) are preserved by the regularization technique and 
they do not have to be restored order by order, and {\it ii)} 
one can provide a mass term for the quantum gauge field 
without destroying the background gauge invariance (whereas 
the STI will be dramatically broken by this choice). These 
important features can be used to pursue the computation 
of the effective action beyond one loop in the present context. However, 
in order to construct the full effective action one has to take into 
account the renormalization of all other identities such as 
(\ref{AGE}), (\ref{NLE}), (\ref{GE}) and (\ref{STI}). 
For what concerns the linear identities, it is easy to 
check that they are indeed preserved by the 
regularization procedure and therefore the algebraic manipulations 
performed in Section 2  remain valid beyond tree level. If one 
of these equations is broken, 
an explicit counterterm given by Sorella and Piguet \cite{book} 
(as a function of the regularization-dependent breaking terms)
is used to restore it order by order, and 
therefore we assume that all equations but (\ref{STI}) are preserved by the 
regularization procedure.  

The Wilson--Polchinski regularization method introduces explicit breaking 
of the STI in the form 
\begin{equation}\label{reno:1}
{\cal S}(\hat \Gamma') = \Delta \cdot \hat \Gamma'  \,, 
\end{equation}
where the R.H.S. denotes the insertion of the breaking term $\Delta$ into 
the Green functions generated by $\hat\Gamma'$. The functional 
dependence of $\Delta$ is restricted by the equations (\ref{NLE}) -- (\ref{WTI}) 
to depend only on background gauge-invariant combinations 
of $A_{\mu}, \hat A_{\mu}, \hat\Omega_{\mu}$ and $c^{*}$.  
In addition, the local approximation of
the breaking term $\Delta$ 
is restricted by power counting to be a sum of local operators with dimension $4$.
On general grounds,  there are additional breakings in the R.H.S.
of equation (\ref{reno:1}), but they are given by irrelevant operators
vanishing in the physical limit of infinite UV cut-off.
It can be proved (see \cite{becchi-lectures} for a complete  
analysis within  gauge theories) 
that at the physical fixed point the STI 
are indeed restored, provided that the breaking terms of dimension $4$
associated with the local approximation of $\Delta$ 
 have been removed
order by order in the loop expansion.

The most general polynomial  $\Delta$ 
with dimension $4$ and ghost-number $1$  is 
\begin{equation}\label{reno:2}
\Delta^{1}_4 = \int d^{4}x \,
\hat \Omega_\mu F^\mu(A, \hat A) 
%+\hat \Omega_\mu \hat \Omega_\nu \hat A^*_\rho G^{\mu\nu\rho}(A,\hat A)
% +\hat \Omega_\mu \hat \Omega_\nu \hat \Omega_\rho c^*
%H^{\mu\nu\rho}(A,\hat A) \,, 
\end{equation}
where $F$ 
%,G$ and $H$ are only 
is a function of $A_{\mu}$ and of $\hat A_{\mu}$ only.
%However, by dimensional reasons $H=G=0$ and we are left with the first term. 
The function $F$ should have dimension 3.
Since we can use integration by 
parts, we assume that $\hat \Omega_{\mu}$ in the first term is 
undifferentiated. 

Notice that 
there is no room in $\Delta^{1}_{4}$ for the representative of the ABJ anomaly. 
However, this is not surprising since if this  anomaly appears, it occurs 
in 
(\ref{WTI}) and (\ref{reno:1}) with the same coefficient. Therefore, 
if we assume that the (\ref{WTI}) are preserved by the regularization, we 
also assume that there is no anomaly in (\ref{reno:1}).
$\Delta^1_4$ is constrained to obey the following Wess--Zumino
consistency condition
\begin{eqnarray}\label{wz_1}
{\cal S}_{\hat \G^{' (0)}} (\Delta^1_4) = 0 \, ,
\end{eqnarray}
where
\begin{eqnarray}\label{wz_2}
{\cal S}_{\hat \G^{' (0)}} = 
\int d^4x \, {\rm tr} \left ( \frac{\delta \hat \G^{' (0)}}{\delta \hat A^*_\mu}
\frac{\delta}{\delta A_\mu} 
+
\frac{\delta \hat \G^{' (0)}}{\delta A_\mu}
\frac{\delta}{\delta \hat A^*_\mu} 
+
\hat \Omega_\mu \frac{\delta}{\delta \hat A_\mu} \right )
\, 
\end{eqnarray}
and $\hat \G^{' (0)}$ is the coefficient of order zero in the $\hbar$-expansion
of the functional $\hat \G^{'}$ (i.e. its classical approximation). 
Furthermore,  ${\cal S}_{\hat \G^{' (0)}}$ is nilpotent.

The variables $(\hat A_\mu, \hat \Omega_\mu)$ have special
transformation properties with respect to ${\cal S}_{\hat \G^{' (0)}}$.
They form a set of coupled doublets, i.e.
\begin{eqnarray}
{\cal S}_{\hat \G^{' (0)}} ( \hat A_\mu ) = \hat \Omega_\mu \, , ~~~~~
{\cal S}_{\hat \G^{' (0)}} ( \hat \Omega_\mu ) = 0 \, ,
\label{doub_1}
\end{eqnarray}
while the counting operator
\begin{eqnarray}
{\cal N} = \int d^4x \, {\rm tr} \left ( \hat A_\mu \frac{\delta}{\delta A_\mu}
+ \hat \Omega_\mu \frac{\delta}{\delta \Omega_\mu} \right ) 
\label{doub_2}
\end{eqnarray}
does not commute with ${\cal S}_{\hat \G^{' (0)}}$.
The cohomological properties of doublet pairs have been
widely discussed in the literature 
\cite{book,Barnich:2000zw,Brandt:2001tg,Brandt:1996mh}.
In particular, coupled doublets have been analysed 
in \cite{Quadri:2002nh}.
It turns out that if $Y$ is a functional belonging to the kernel
of ${\cal S}_{\hat \G^{' (0)}}$ and that $\left . Y \right |_{\hat \Omega=\hat A=0} = 0$, then it can be expressed as a ${\cal S}_{\hat \G^{' (0)}}$-variation of some suitable functional $X$, so that $Y =  {\cal S}_{\hat \G^{' (0)}}(X)$.
Since $\Delta^1_4$ is in the kernel of ${\cal S}_{\hat \G^{' (0)}}$
and vanishes at $\hat \Omega=0$,  it can be written as
\begin{eqnarray}
\Delta^1_4 = {\cal S}_{\hat \G^{' (0)}}(\Xi[A,\hat A])
\label{doub_3}
\end{eqnarray}
for some local counterterm $\Xi[A,\hat A]$ of dimension $4$,  
which reabsorbs the breaking.
As a consequence of the local antighost equation the cohomology
associated to the STI has been trivialized. This property
is a distinctive feature of the Landau BFM gauge.
It does not hold in the ordinary Landau gauge. A detailed
analysis of the structure of the counterterms in Yang--Mills theory
in the ordinary Landau gauge has been given in \cite{quadri:landau}.
%
%%%%%%%%%%%%%%%%%%%%%%%%%%%%%%%%%%%%%%%%%%%%%%%%%%%%%%%%%%%%

The background gauge invariance also imposes restrictions on 
the number of independent polynomials that can appear in $F_\mu$. 
Indeed by imposing background gauge invariance we find that $F_\mu$
must have the form (here 
we use the notation $Q_{\mu} = A_{\mu} - \hat A_{\mu}$ for 
simplicity) 
\begin{eqnarray}\label{reno:3}
F^{\mu}_d &=& 
m^{\mu}_{dabc, \nu\rho\sigma}
Q^a_\nu Q^b_\rho Q^c_\sigma +  
n^{\mu}_{dab, \nu\rho\sigma}
Q^a_\nu (\hat\nabla_\rho Q_\sigma)^b + 				      
p^{\mu}_{dab, \nu\rho\sigma}
Q^a_\nu \hat F^b_{\rho \sigma} + \nonumber \\
&+& p^{\mu}_{da, \nu\rho\sigma}
\hat\nabla_{(\nu} \hat\nabla_{\rho)} Q^a_\sigma	+ 
r^{\mu}_{da, \nu\rho\sigma}	      
(\hat\nabla_\nu \hat F_{\rho\sigma})^a \,, 				     
\end{eqnarray}
where $m,n,p,q$ and $r$ are constant Lorentz- and Lie algebra tensors 
(independent of the fields and their derivatives).
They
have to be computed directly from the Feynman diagrams in a perturbative 
expansion or in terms of the effective action for non-perturbative evaluation. They 
depend on the regulator and they are finite quantities at the fixed point. 
Once $F^\mu_d$ is known,  the functional $\Xi$ in equation (\ref{doub_3})
can be explicitly obtained
by making use of the inversion formulae given in \cite{Quadri:2002nh}.

The BFM of the 
Landau gauge permits such enormous simplifications to all orders,   
reducing the effort to restore the STI. Notice that in the absence of a 
good regularization procedure (for example for chiral gauge 
theory on the continuum), one can always find the regularization 
 that  preserves the WTI identities and few breaking 
terms of the STI have to be computed order by order.

We wish to conclude this section by stressing that 
the combination of Landau gauge fixing, covariantized with respect 
to the background fields, provides a way of  significantly
simplifying  the renormalization of the STI by trivializing the
cohomology of the relevant linearized ST operator in equation (\ref{wz_2}).
In the presence of a background gauge-invariant regularization 
procedure,  such as the regularization adopted in 
\cite{bonini}, the number of breaking terms (and consequently the 
number of counterterms) of the STI is henceforth highly reduced,
resulting in a much simpler construction of the quantum
effective action.

%%%%%%%%%%%%%%%%%%%%%%%%%%%%%%%%%%%%

%%%%%%%%%%%%%%%%%%%%%%%%%%%%%%%%%%%%

\subsection{Fixed points and relation between the IR behaviour of 
$E^{T}(p)$ and $C^{L}(p)$}

Following Ref.~\cite{Lerche:2002ep}, we derive the relation between 
the two-point functions $E^{T}(p)$ and $C^{L}(p)$ by assuming 
only the existence of a fixed point. The present derivation is based 
on the properties of the Landau-BFM. 

From the STI (\ref{STI}), by differentiation with respect to 
$A_{\mu}, A_{\nu}$ and $c$, we have 
\begin{eqnarray}\label{fpI}
&&\Gamma_{c A^{* \rho}}(-p-q) \Gamma_{A_{\rho} A_{\mu} A_{\nu}}(p,q) +
\nonumber \\
&& 
+\Gamma_{c A^{* \rho} A_{\mu}}(-q,p) \Gamma_{A_{\rho} A_{\nu}}(q) + 
\Gamma_{c A^{* \rho} A_{\nu}}(-p,q) \Gamma_{A_{\rho} A_{\mu}}(p) = 0\,.
\end{eqnarray}
The Green function $\Gamma_{c A^{* \rho} A_{\mu}}(-q,p)$ 
is finite as shown in the previous section. By using the transversality 
of $ \Gamma_{A_{\rho} A_{\nu}}(q)$ and the Lorentz decomposition 
of $\Gamma_{c A^{* \rho}}$ given in (\ref{II}), we get
\begin{eqnarray}\label{fpII}
&& \!\!\!\!\!\!\!\!\!\!\!\!\!\! i (C^{L}(-p-q) +1) 
\left[(-p-q)_{\rho} \Gamma_{A_{\rho} A_{\mu} A_{\nu}}(p,q) \right] +
\nonumber \\
&& 
\!\!\!\!\!\!\!\!\!\!\!\!\!\! +
\left[\Gamma_{c A^{* \rho} A_{\mu}}(-q,p) T_{\rho \nu}(q)\right] E^{T}(q) +
\left[\Gamma_{c A^{* \rho} A_{\nu}}(-p,q) T_{\rho \mu}(p)\right] E^{T}(p)
= 0\,.
\end{eqnarray}
By contracting the above equation with the tensor
$p^{\mu} p^{\nu}/p^{2}$,  the 
third term vanishes because it is transverse, and 
we have 
\begin{eqnarray}\label{fpIII}
(C^{L}(-p-q;\mu) +1) S(p,q;\mu) + 
R(p,q;\mu) E^{T}(q;\mu) = 0\,,
\end{eqnarray}
where we have defined the following 
scalar functions
\begin{eqnarray}
S(p,q; \mu) &=&{i \over p^{2}}  
\left[(-p-q)_{\rho} \Gamma_{A_{\rho} A_{\mu} A_{\nu}}(p,q; \mu) 
p^{\mu} p^{\nu} \right] 
\nonumber \\
R(p,q; \mu) &=& {1\over p^{2}}
\left[\Gamma_{c A^{* \rho} A_{\mu}}(-q,p; 
\mu) T_{\rho \nu}(q) p^{\mu} p^{\nu}\right]
\end{eqnarray}
and $\mu$ is the renormalization scale. 
 Since the function $S(p,q; \mu)$ is finite, its 
 variation under a variation of the renormalization scale $\mu$ can be 
 given entirely in terms of the functions $\partial_{\mu} E^{T}(p;\mu)$ 
 and $\partial_{\mu} C^{L}(p;\mu)$:
\begin{equation}\label{fpIV}
\partial_{\mu} S(\mu^{*}; \mu) = 
X_{E}(\mu^{*};\mu) \partial_{\mu} E^{T}(\mu^{*};\mu) +
X_{C}(\mu^{*};\mu) \partial_{\mu} C^{L}(\mu^{*};\mu)
\end{equation}
 computed at the symmetric point $p^{2}=q^{2}=(\mu^{*})^{2}$. 
Assuming the existence of a fixed point for the 
three-point function $R(\mu^{*};\mu) =0$ at a given scale $\mu^{*}$ 
%--- 
% the existence of a fixed point can be deduced by the 
%existence of a fixed point for the gauge coupling  ---  
we find that 
\begin{equation}
\label{fpV}
\partial_{\mu} E^{T}(\mu^{*};\mu)  Y_{E} + 
\partial_{\mu} C^{L}(\mu^{*};\mu) Y_{C} =0\,, 
\end{equation}
which relates the $\mu$-dependence of the 
two-point functions. The functions $Y_{E}$ and $Y_{C}$ 
are finite and completely fixed by the normalization conditions. The 
existence of a fixed point for the three point function 
$R(\mu^{*};\mu) =0$ can be deduced from the existence 
of a fixed point for the running of beta function for the gauge coupling. 
The difference between the two fixed points amounts to a change of 
the renormalization prescriptions for the two-point functions.

%%%%%%%%%%%%%%%%%%%%%%%%%%%%%%%%%%%%%%

\subsection{Kugo--Ojima criterion on  confinement using BFM}

A thorough discussion of the Kugo--Ojima (KO) criterion on  confinement 
is beyond the scope of this note and we refer the reader to the original 
literature \cite{Kugo:gm}. However, before analysing the criterion 
within the  Landau BFM,  some critical remarks are in order.  
First of all, one should mention that the KO criterion  is established within 
the asymptotic Fock space and is  therefore  intimately related with 
perturbation theory. Moreover, in contrast to the gauge-invariant  
Wilson-loop criterium of confinement,  the KO criterion is established 
in the gauge-fixed theory and has to face the serious and well-known 
problem of  Gribov copies. 

Keeping these problems in mind,  we now  translate 
the KO criterion  for the confinement 
into  the  context  of the Landau background gauge. The advantage of this
framework is  that all composite operators needed in the 
present analysis are already present in the Lagrangian (\ref{lag}) coupled 
to the sources $\Omega_{\mu}^{a}$ and $A^{*,a}_{\mu}$. 

First we compute the Noether current for the $SU(N)$ invariance of the 
action (\ref{lag}). Simple algebraic manipulations 
lead to 
\begin{eqnarray}\label{KOI}
- i J_{\mu} = [A^{\nu}, F_{\mu\nu}] +  [b, (A-\hat A)_{\mu}] 
+ [\bar c, \Omega_{\mu} ]- [\bar c, \nabla_{\mu} c]  - [\hat\nabla_{\mu} \bar c,c] \,,
\end{eqnarray}
which is conserved (using the equations of motions for the classical fields). 

If we compute the equation of motion for the gauge field $A_{\mu}$, 
we obtain a related expression
\begin{eqnarray}
\label{KOII}
\frac{\delta S}{\delta A_{\mu}} &=& \nabla^{\nu} F_{\nu\mu} - \partial_{\mu}\, b 
- i [b, \hat A_{\mu}] + i [\nabla \bar c, c] + i [\bar c, \Omega]  \nonumber \\
&=& \partial^{\nu} F_{\nu\mu} - J_{\mu} - s\, \left[ \nabla_{\mu} \bar c \right] \nonumber\\
&=& \partial^{\nu} F_{\nu\mu} - J_{\mu} + \frac{\delta S}{\delta \hat A_{\mu}} \,,  
\end{eqnarray} 
where we used in the last line the property that, in terms of $A_{\mu}$ and 
$\hat A_{\mu}$, the Lagrangian (\ref{lag}) depends on the background 
only through the gauge fixing. 

We neglect the contribution of the antifields since they drop 
out from the computation of the previous current on the 
Fock space of on-shell states. In addition, for on-shell states
one knows that the L.H.S. of (\ref{KOII}) vanishes.
So the 
KO charges can be defined by 
\begin{equation}\label{KOIII}
G = \int d^{3}x \, \partial^{i} F_{i 0}\,, ~~~~~~ 
N = \int d^{3}x  \, \frac{\delta S}{\delta \hat A_{0}} \,, ~~~~
Q = \int d^{3}x \, J_{0}\,,
\end{equation}
which are related by $Q = N +G$. The KO criterion  asserts that 
if there is a mass gap in the gluon propagator $\lim_{p^{2} \rightarrow 0} p^2 E^{T}(p)  
= 0$, the first charge $G$ vanishes because it 
is a total derivative. The second condition 
imposes that also $N$ should vanish, so that $Q$ is well-defined and all  
coloured states vanish on physical states. Notice that the second condition can  expressed by studying the correlation function between the current ${\delta S} /{\delta \hat A_{\mu}}$ 
and a one-particle state with a gluon field, namely it can be related
to the two-point function ${\delta^2 S}/{\delta \hat A_\mu(x) A_{\nu}(y)}$
  of the background and quantum gauge field. 

%%%%%%%%%%%%%%%%%%%%%%%%%%%%%%%%%%%%%%%

\subsection{The correlation function 
$\langle T \nabla^\mu c(x) \nabla^\nu \bar c(y) \rangle$}

The KO criterion can be reformulated 
in terms of the IR behaviour of a special correlation function
involving the ghost fields.
More precisely, one defines the matrix $u^{ab}(p^2)$ as

\begin{figure}
\begin{center}
\epsfig{file=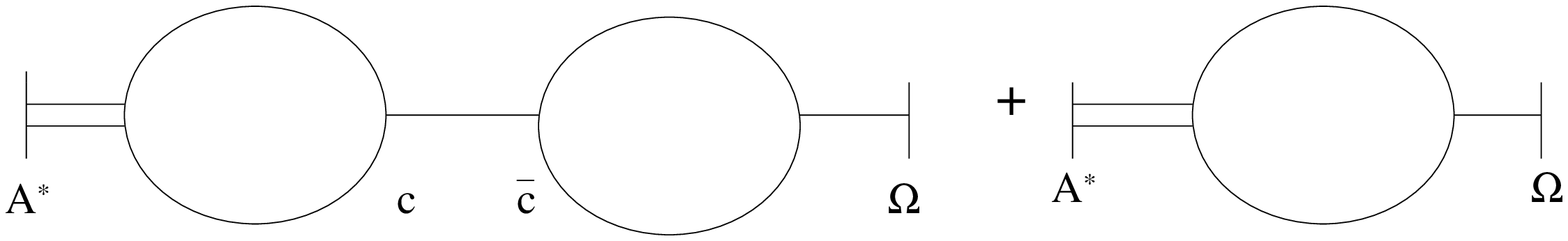, width= 10 cm}
\caption{Contribution to ${\cal G}_{\mu\nu}$ from its connected components}
\label{fig1}
\end{center}
\end{figure}
\begin{eqnarray}
\int d^4x e^{i p (x-y)} \langle (\nabla_\mu c)^a (x) (f^{bcd} A_\nu^c \bar c^d)(y) \rangle
=:_{p \rightarrow 0} T_{\mu\nu} u^{ab}(p^2) \, , 
\label{corr_1}
\end{eqnarray}
where the equality is valid in the limit of small momenta (i.e. one
discards possible contributions from massive states to the correlation
function in the L.H.S. of equation (\ref{corr_1})).
Then, according to Kugo and Ojima, the charge $N$ is well defined provided
that $u^{ab}(p^2)$ fulfills the condition \cite{Alkofer}
\begin{eqnarray}
u^{ab}(0) = -\delta^{ab} \, .
\label{corr_2}
\end{eqnarray}
In the Landau background gauge, the Green function 
$\langle T \nabla^\mu c(x) \nabla^\nu \bar c(y) \rangle$ can be generated
by functional differentiation with respect to the antifield
$A^*_\mu$ and the background ghost $\Omega_\nu$:
\begin{eqnarray}
{\cal G}_{\mu\nu}(x-y) = \langle T \nabla^\mu c(x) \nabla^\nu \bar c(y) \rangle^C = \left . 
\frac{\delta^2  W}{\delta A^*_\mu(x) \delta \Omega_\nu(y)} 
\right |_{J=\beta=0} \, .
\label{dse_1}
\end{eqnarray}
In the Landau background field gauge many simplifications arise
for the computation of ${\cal G}_{\mu\nu}$.
First we observe that there are two contributions to ${\cal G}_{\mu\nu}$
coming from the connected graphs, which are shown in Figure \ref{fig1}.

Analytically we obtain, in the momentum space, using the
parameterization in equation (\ref{II}):
\begin{eqnarray}
- {\cal G}_{\mu\nu}(p) & = & - p_\mu A(p) \frac{G(p)}{p^2} p_\nu B(p) + 
                         L_{\mu\nu}C^L(p) + T_{\mu\nu} C^T(p) \nonumber \\
                       & = & -G^{-1}(p) L_{\mu\nu} + 
                             (G^{-1}(p) - 1) L_{\mu\nu} + C^T(p) T_{\mu\nu}
                         \nonumber \\
                       & = & - L_{\mu\nu} + C^T(p) T_{\mu\nu} \, .
\label{dse_2}
\end{eqnarray}

\unitlength=1bp
\newcount\boxwidth
\boxwidth=100
\newdimen\boxlength
\boxlength=\boxwidth\unitlength
\begin{small}
\begin{figure}
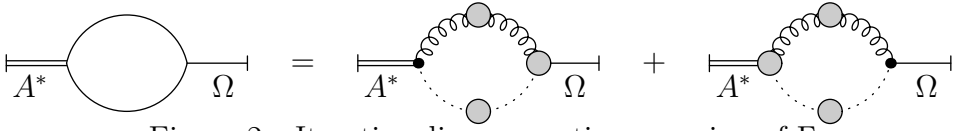

\begin{feynartspicture}(\boxwidth,\boxwidth)(1,1)
\FADiagram{}
\FAProp(0.,10.2)(5.,10.2)(0.,){/Straight}{0}
\FAProp(0.,9.8)(5.,9.8)(0.,){/Straight}{0}
\FAProp(0.,9.3)(0.,10.7)(0.,){/Straight}{0}  % vertical left
\FALabel(2.,9)[t]{$A^*$}
\FAProp(20.,10.)(15.,10.)(0.,){/Straight}{0}
\FAProp(20.,9.4)(20.,10.6)(0.,){/Straight}{0}  % vertical right
\FALabel(18.,9)[t]{$\Omega$}
\FAProp(5.,10.)(15.,10.)(0.8,){/Straight}{0}
%\FALabel(10.,5.98)[t]{$3$}
\FAProp(5.,10.)(15.,10.)(-0.8,){/Straight}{0}
%\FALabel(10.,14.02)[b]{$4$}
%\FAVert(5.,10.){0}
%\FAVert(15.,10.){-1}
%\FAVert(10.,6.){-1}
%\FAVert(10.,14.){-1}
\end{feynartspicture}
~~\raise .47\boxlength\hbox{=}~~
\begin{feynartspicture}(\boxwidth,\boxwidth)(1,1)
\FADiagram{}
\FAProp(0.,10.2)(5.,10.2)(0.,){/Straight}{0}
\FAProp(0.,9.8)(5.,9.8)(0.,){/Straight}{0}
\FAProp(0.,9.3)(0.,10.7)(0.,){/Straight}{0}  % vertical left
\FALabel(2.,9)[t]{$A^*$}
\FAProp(20.,10.)(15.,10.)(0.,){/Straight}{0}
\FAProp(20.,9.4)(20.,10.6)(0.,){/Straight}{0}  % vertical right
\FALabel(18.,9)[t]{$\Omega$}
\FAProp(5.,10.)(15.,10.)(0.8,){/GhostDash}{0}
\FAProp(5.,10.)(10.,14.)(-0.4,){/Cycles}{0}
\FAProp(10,14.)(15.,10.)(-0.4,){/Cycles}{0}
\FAVert(5.,10.){0}
\FAVert(15.,10.){-1}
\FAVert(10.,6.){-1}
\FAVert(10.,14.){-1}
\end{feynartspicture}
~~\raise .47\boxlength\hbox{+}~~
\begin{feynartspicture}(\boxwidth,\boxwidth)(1,1)
\FADiagram{}
\FAProp(0.,10.2)(5.,10.2)(0.,){/Straight}{0}
\FAProp(0.,9.8)(5.,9.8)(0.,){/Straight}{0}
\FAProp(0.,9.3)(0.,10.7)(0.,){/Straight}{0}  % vertical left
\FALabel(2.,9)[t]{$A^*$}
\FAProp(20.,10.)(15.,10.)(0.,){/Straight}{0}
\FAProp(20.,9.4)(20.,10.6)(0.,){/Straight}{0}  % vertical right
\FALabel(18.,9)[t]{$\Omega$}
\FAProp(5.,10.)(15.,10.)(0.8,){/GhostDash}{0}
\FAProp(5.,10.)(10.,14.)(-0.4,){/Cycles}{0}
\FAProp(10,14.)(15.,10.)(-0.4,){/Cycles}{0}
\FAVert(5.,10.){-1}
\FAVert(15.,10.){0}
\FAVert(10.,6.){-1}
\FAVert(10.,14.){-1}
\end{feynartspicture}
\vspace{-1.5cm}
\caption{\label{fig:2} Iterative diagrammatic expansion of $\G_{A^*\Omega}$.}
\end{figure}
\end{small}

In the above equation we have used the simplifications coming
from equation (\ref{ III}).
This means that in the Landau background gauge 
the longitudinal part of ${\cal G}_{\mu\nu}$ does not
get renormalized and the only dynamical information is contained
in $C^T(p)$. Moreover, this parameter can be extracted by looking
at the 1-PI Green function $\G_{\Omega_\mu A^*_\nu}$:
\begin{eqnarray}
C^T(p) = - T^{\mu\nu} \G_{\Omega_\mu A^*_\nu} \, .
\label{dse_3}
\end{eqnarray}
By comparing equation (\ref{corr_1}) with equation (\ref{dse_1}) we arrive at the following
identification:
\begin{eqnarray}
C^T(p^2) = -u(p^2) \, .
\label{dse_4}
\end{eqnarray}
Therefore the full information contained in the matrix $u$ can actually
be derived from the knowledge of $\G_{\Omega_\mu A^*_\nu}$
via equation (\ref{dse_3}). 

\medskip
In order to compute the 1-PI Green function $\G_{A^*\Omega}$, 
we propose an iterative diagrammatic prescription motivated
by the perturbative treatment and by some analogy with
the SDE. 
Graphically we can represent it as in Figure \ref{fig:2}. 
Notice that $\G^{(0)}_{A^*\Omega}=0$. Then we can write
\begin{eqnarray}
\!\!\!\!\!\!
\G_{A^{a*}_\mu(-p) \Omega^b_\nu(p)} & =& \int d^4q 
\G^{(0)}_{A^{*a}_\mu(-p) A^j_\nu(q) c^c(p+q)} 
D^{\nu\rho}_{ji}(q) \G_{A^i_\rho(q) \Omega^b_\nu(p) \bar c^k(p+q)}
D^{kc}_{c \bar c}(p+q) + \nonumber \\
&& \!\!\!\!\!\!\!\!\!\!\!\! \int d^4q 
\G_{A^{*a}_\mu(-p) A^j_\nu(q) c^c(p+q)} 
D^{\nu\rho}_{ji}(q) \G^{(0)}_{A^i_\rho(q) \Omega^b_\nu(p) \bar c^k(p+q)}
D^{kc}_{c \bar c}(p+q)\,. 
%\nonumber \\
\label{sde}
\end{eqnarray}
Here, $D^{\nu\rho}_{ji}$ and $D^{kc}_{c \bar c}$ stand for
the full propagators of the gluons and the ghosts, 
$\G^{(0)}_{A^{*a}_\mu A^j_\nu c^c}$ is the 
antifield-gluon-ghost tree-level vertex
and 
$\G_{A^{*a}_\mu A^j_\nu c^c}$ the full renormalized one,
and analogously for $\G^{(0)}_{A^i_\rho \Omega^b_\nu \bar c^k}$
and $\G_{A^i_\rho \Omega^b_\nu \bar c^k}$.

Equation (\ref{sde}) could be used as a starting point to derive, by iterative 
solutions, the information
on $C^T(p)$ in equation (\ref{dse_3}), possibly also beyond
standard perturbation theory.

%%%%%%%%%%%%%%%%%%%%%%%%%%%%%%%%%%%%%%%

\section{Conclusion and Outlook}\label{sec:concl}

In the present paper, the complete algebraic structure 
of the BRST symmetry for Landau background gauge fixing 
is exploited. We recover all the results presented in the 
literature for the non-renormalization of the 
trilinear ghost-gluon vertex and different simplifications provided 
by the BFM. In addition, we show that all the information regarding the 
IR behaviour of the theory is encoded in a new Green function 
$\Gamma_{\Omega_{\mu} A^{*}_{\nu}}(p)$ which appears naturally in the 
Landau background gauge.  We provide a prescription 
to compute this new Green function to all orders in perturbation 
theory and beyond the perturbative analysis.  

It may be hoped that the present formalism could enhance 
the distinctive simplifications of the Landau gauge, which is
widely used in the study of
the IR properties of Green functions for QCD and 
in the strong regime of gauge theories.

%%%%%%%%%%%%%%%%%%%%%%%%%%%%%%%%%%%%

\section*{Acknowledgements}

P.A.G. and A.Q.  thank CERN for the hospitality and the 
financial support. 
We acknowledge illuminating discussions with M. L\"uscher, R. Stora, 
Ph. de Forcrand, R. Sommer, D. Litim and J.M. Pawlowski. 
P.A.G. thanks G. Sterman for the invitation to lecture 
at YITP (Stony Brook) on the BFM which  initiated the present          
paper.  

%%%%%%%%%%%%%%%%%%%%%%%%%%%%%%%%%%%%%

\end{document}